\documentclass{aa}
\usepackage{graphicx}
\usepackage{rotating}
\def\gtrsim{\mathrel{\hbox{\rlap{\hbox{\lower4pt\hbox{$\sim$}}}\hbox{$>$}}}}
\def\ltsim{\mathrel{\hbox{\rlap{\hbox{\lower4pt\hbox{$\sim$}}}\hbox{$<$}}}}

\def\kms{\hbox{{\rm km}\,{\rm s}$^{-1}$}}
\def\bz{\hbox{$\langle B_z\rangle$}}

 \def\ii{\,{\sc ii}} \def\iii{\,{\sc iii}}

\begin{document}
   \title{The magnetic Bp star 36 Lyncis\\I. Magnetic and photospheric properties\thanks{Based on observations obtained using the MuSiCoS spectropolarimeter at Pic du Midi Observatory, the Canada-France-Hawaii Telescope (operated by the National Research Council of Canada, the Centre National de la Recherche Scientifique of France, and the University of Hawaii), the Dominion Astrophysical Observatory, Herzberg Institute of Astrophysics, National Research Council of Canada, the David Dunlap Observatory, University of Toronto, and the Elginfield Observatory, University of Western Ontario.}}

   \author{G.A. Wade\inst{1}, M.A. Smith\inst{2}, D.A. Bohlender\inst{3}, T.A. Ryabchikova\inst{4,6}, 
   C.T. Bolton\inst{5}, \\T. Lueftinger\inst{4}, J.D. Landstreet\inst{7}, P. Petit\inst{8}, S. Strasser\inst{9}, M. Blake\inst{5}, G.M. Hill\inst{10}}

   \offprints{G.A. Wade}

   \institute{Department of Physics, Royal Military College of Canada, 
                 PO Box 17000, Station 'Forces', Kingston, Ontario, Canada K7K 4B4
   \and
   Computer Sciences Corporation/Space Telescope Science Institute, 
                 3700 San Martin Dr., Baltimore, MD, 21218, USA
   \and
   National Research Council of Canada, Herzberg Institute of Astrophysics, 
                 5071 West Saanich Road, Victoria, BC, Canada V9E 2E7
   \and
    Institute f\"ur Astronomie, Universitat Wien, T\"urkenschanzstrasse 17,
    1180 Vienna, Austria
   \and
   David Dunlap Observatory, University of Toronto, P.O. Box 360, Richmond Hill, ON, Canada, L4C 4Y6 
   \and
   Institute of Astronomy, Russian Academy of Sciences, Pyatnitskaya 48, 109017 Moscow, Russia
   \and
   Department of Physics \& Astronomy, The University of Western Ontario, London, Ontario, Canada, N6A 3K7
   \and
          Max-Planck Institut f\"ur Aeronomie
           Max-Planck-Str. 2
           37191 Katlenburg-Lindau, Germany
   \and
   Department of Astronomy, University of Minnesota, 116 Church Street, S.E., Minneapolis, Minnesota 55455
   \and
   W. M. Keck Observatory, 65-1120 Mamalahoa Highway, Kamuela, HI, USA, 96743}           

   \date{Received ??; accepted ??}


   \abstract{}{This paper reports the
photospheric, magnetic and circumstellar gas characteristics of the magnetic B8p
star 36 Lyncis (HD 79158).}{Using archival data and new polarised and unpolarised high-resolution spectra, we 
redetermine the basic physical properties, the rotational period and the geometry of
the magnetic field, and the photospheric abundances of various
elements.}{Based on magnetic and spectroscopic measurements, we infer an improved rotational period of $3.83475\pm 0.00002$~d. We determine a current epoch of the longitudinal magnetic field positive extremum (HJD 2452246.033), and provide constraints on the geometry of the dipole magnetic field ($i\geq 56\degr$, $3210~{\rm G}\leq B_{\rm d}\leq 3930$~G, $\beta$ unconstrained). We redetermine the effective temperature and surface gravity using the optical and UV energy distributions, optical photometry and Balmer line profiles ($T_{\rm eff}=13300\pm 300$~K, $\log g=3.7-4.2$), and based on the Hipparcos parallax we redetermine the luminosity, mass, radius and true rotational speed ($L=2.54\pm 0.16~L_\odot, M=4.0\pm 0.2~M_\odot, R=3.4\pm 0.7~R_\odot, v_{\rm eq}=45-61.5$~\kms). We measure photospheric abundances for 21 elements using optical and UV spectra, and constrain the presence of vertical stratification of these elements. We perform preliminary {preliminary} Doppler Imaging of the surface distribution of Fe, finding that Fe is distributed in a patchy belt near the rotational equator. Most remarkably, we confirm strong variations of the H$\alpha$ line core which we interpret as due to occultations of the star by magnetically-confined circumstellar gas.}{}{}
   \maketitle

\section{Introduction}

The magnetic Bp stars exhibit a rich variety of magnetic and wind phenomena, many of which are unique to this class of star. Landstreet \& Borra (1978), detecting for the first time a magnetic field in a helium strong star, proposed that the photometric and spectroscopic variations of $\sigma$~Ori E resulted from hot gas trapped above the magnetic equator. Brown et al. (1985) discovered unexpectedly strong C~{\sc iv}, Si~{\sc iv} and N~{\sc v} ultraviolet
resonance line absorption and emission that varied with the rotational period of the magnetic helium weak 
star HD 21699. Subsequent work revealed similar phenomena in a
number of other Bp stars, which was attributed to rotational modulation of magnetically-structured stellar winds (e.g. Shore 1987, Shore et al. 1990, Shore \& Brown 1990, Bolton 1994,
Donati et al. 2001, Smith \& Groote 2001, Neiner et al. 2003). The
additional discovery of non-thermal radio emission (Drake et al. 1987)
from a subset of these objects resulted in the proposal that these
stars are surrounded by an extended wind-generated magnetosphere
(Linsky et al. 1992), in which the radio luminosity results from
optically thick gyrosynchrotron emission and should therefore 
correlate with both magnetic field strength and effective
temperature. According to Drake et al. { (2002)}, by 2002 about 120
magnetic CP stars had been observed in the radio, and of these about
25\% had been confidently detected. The general existence of this phenomenon,
and the establishment of correlations between effective temperature, magnetic
field strength, rotational period and radio emission suggest that this phenomenon
is intrinsic to Ap and Bp stars, and is related to their magnetic properties.
Subsequently, X-ray emission was
discovered (Drake et al. 1994) from a small number of these objects. As pointed
out by Drake et al. { (2002)}, no clear correlation of X-ray brightness
with other physical or observational characteristics (including radio
luminosity) has been established, suggesting (at least in the A-type magnetic stars)
that X-ray emission may be due to the presence of late-type companions, and unrelated
to their magnetic properties\footnote{There exists compelling evidence, in the form of variability of the X-ray flux according to the stellar rotational period, that some magnetic B and O stars (the Bp star AB Aur: Babel \& Montmerle (1997); the magnetic O7V star $\theta^1$~Ori C: Gagn\'e et al. (1997), Donati et al. 2002) are intrinsic X-ray emitters. For the majority of magnetic upper-main sequence stars, however, this is not the case.}. More recently, sophisticated models 
describing the hydrodynamic interaction between the magnetic field and
wind flow have been developed (Babel \& Montmerle 1997, ud-Doula \&
Owocki 2002, Preuss et al. 2004, Townsend \& Owocki 2005), leading to the possibility of combining the various
individual kinds of observations (magnetic, radio, optical, UV, X-ray)
into a  single coherent picture of the wind-magnetic field interaction
and subsequent circumstellar structure of magnetic Bp stars.

36 Lyncis (HD 79158) is a bright ($V=5.3$), well-studied magnetic
(Borra, Landstreet \& Thompson 1983) helium weak B star. It is
classified as B8IIIpMn in the Bright Star Catalogue (Hoffleit et
al. 1995), as a Si star by Molnar (1972), as a Mn star by Cowley
(1972), and as a Sr Ti helium weak star by Borra \& Landstreet (1983). It was first identified as peculiar by Edwards (1932). The
spectrum and abundances have been studied by Searle \& Sargent (1964),
Mihalas \& Henshaw (1966) and Sargent, Greenstein \& Sargent
(1969). According to these studies, lines of neutral He and O are
weak, whereas those of C, Ne, Si, Mg and Cr are only marginally
anomalous. Lines of Ti and Fe are anomalously strong. HD 79158 is a
member of the {\em sn} class (Abt \& Levato 1978), which means its Si~{\sc ii} lines are sharp, whereas He~{\sc i} lines are nebulous. Ryabchikova \& Stateva (1995) { concluded} that 36 Lyn is a classical helium weak star.

Using IUE spectra, Sadakane (1984) reported the presence of strong
lines coincident with the position of the C~{\sc iv} $\lambda\lambda
1548, 1550$ and Si~{\sc iv} $\lambda\lambda 1394, 1403$ resonance
doublets. Shore, Brown \& Sonneborn (1987) obtained additional
magnetic field measurements and IUE spectra of 36 Lyn and two
additional {\em sn} stars, confirming the presence of the
``superionised'' C~{\sc iv} and Si~{\sc iv} resonance lines and
detecting variability of these features. Shore et al. (1990) extended
these observations and analysis, obtaining 22 new measurements of the
longitudinal magnetic field and determining a rotational period of
$3.8345\pm 0.001$ days.

The goal of the present paper is to provide an accurate reassessment
of the magnetic and photospheric properties of 36 Lyn as the basis for
a detailed study of properties of the circumstellar matter (Smith et
al. 2006, ``Paper II''). 

\section{New and archival observations}

\subsection{Unpolarised optical spectra}

 The journal of the unpolarised optical spectroscopic observations is presented in Table 1, available only in electronic form from the CDS.

\subsubsection{DAO and DDO Spectra}

H$\alpha$ observations of 36\,Lyn were obtained at the Dominion Astrophysical
Observatory (DAO) with both the 1.85\,m Plaskett (1991, 1992, and 1997-2000) 
and 1.22\,m McKellar (2001, 2003-2005) telescopes, and at the David Dunlap Observatory 1.88\,m telescope (1992, 1997, and 2004).

At DAO, on the 1.85\,m we used the Cassegrain spectrograph 
with the 21\,inch $f/5$ camera,  1800\,line\,mm$^{-1}$ grating in first order, 
VSIS21B image slicer, and various CCDs.
The resulting spectra covered a minimum of approximately 100~\AA\ and, when 
possible, included the \ion{He}{i} $\lambda$6678 feature. 
The dispersion of 10~\AA\,mm$^{-1}$ and effective slit width of 40\,$\mu$m gave a  resolution of approximately $R = 16400$. 

The 2001 McKellar observations were obtained on the coud\'{e} spectrograph with the short camera, 1200\,line\,mm$^{-1}$ holographic grating in first order,
IS32R image slicer, and the UBC-1 CCD.
These short camera observations have a dispersion of 10.1~\AA\,mm$^{-1}$ and
the resolution is approximately $R = 20000$.

For the 2003, 2004 and 2005 McKellar data we again used the coud\'{e} spectrograph, 
but with the long camera, 830\,line\,mm$^{-1}$ grating in first order,
IS32R image slicer, and the SITe4 CCD with 15\,$\mu$m pixels.
The dispersion of 4.8~\AA\,mm$^{-1}$ and projected slit width of 36\,$\mu$m 
produced a resolution of $R = 38000$.

Exposure times on both telescopes ranged from 10 to 30 minutes. 

DDO spectra obtained in 1992 and 1997 were recorded with a Thomson 1024$\times$1024 CCD on the
Cassegrain spectrograph of the 1.88\,m telescope.  
The effective resolution 
was ${\rm R} = 11600$.  
Most of the spectra were centered near 6600~\AA\  so that the 200~\AA\  range covered by the detector included both H$\alpha$ and the He~{\sc i} $\lambda$6678~\AA\  line.

  The DDO spectra obtained in 2004 were recorded with a Jobin-Yvon
800$\times$2000 thinned, back illuminated CCD with 15\,$\mu$m pixels. The 
effective resolution of these spectra is 14700.  These observations were 
taken in the same way as the earlier observations.  However, there
were difficulties because the detector system was not operating properly.
Due to the large RFI introduced by the heater system and the lack of a cold
finger in the dewar, we had to operate the detector system with the temperature
floating near the liquid nitrogen vaporization temperature.  As a result, there
were large changes in the CCD temperature as a function of hour angle of the
telescope.  To minimize this problem, we took frequent biases, and linearly
interpolated in time between them to obtain the bias for each stellar spectrum. The low operating temperature also reduced the dynamic range of the CCD.  Flat
field exposures showed that the response started to become nonlinear at 24k
e$^-$.  As a result, these spectra have somewhat lower S/N than our earlier
spectra.

All data were reduced with  
IRAF\footnote{IRAF is distributed by the National Optical Astronomy 
Observatory, which is operated by the Association of Universities for 
Research in Astronomy (AURA), Inc., under cooperative agreement with the 
National Science Foundation.} in a conventional manner, including bias
subtraction, flat-fielding, wavelength calibration, continuum normalisation
and removal of telluric absorption lines.

\subsubsection{CFHT Spectra}

Canada-France-Hawaii Telescope (CFHT) data were obtained on 
1991 November 19-22 and 1995 October 3 with the coud\'{e} spectrograph and red  
mirror train, $f/8.2$ spectrograph, 600\,line\,mm$^{-1}$ grating in first  
order, and red Richardson image slicer. 
Detectors for these runs were the 1872 diode Reticon array (1991) and Loral3  
(1995) CCD.
The effective resolution was approximately $R = 22000$. 
Exposure times were approximately 15 minutes with the Reticon and 10 minutes 
with the CCD. 
A thorium-neon hollow cathode lamp was used as a comparison line source for 
the Reticon data while a thorium-argon lamp was used for the CCD observation. 
In the case of the Reticon data, each spectrum (including stellar, comparison  
lamp, and flat-field exposures) consisted of a raw exposure and eight 1 second  
baseline exposures and the first step of the data reduction was the subtraction 
of the average of these baselines from each raw spectrum. 
Flat-fielding, wavelength calibration, continuum rectification, and telluric 
line removal was then carried out as described for the DAO observations above. 
The CFHT CCD observations were also reduced in an identical fashion to the DAO  
data. 
 
Heliocentric velocity corrections have been applied to all of the unpolarised spectra. 

\setcounter{table}{1}

\subsection{Circular and linear polarisation spectra}

 
\begin{table*}[t]
\begin{tabular}{ccc|ccc|ccc|ccc} 
\hline
UT Date & HJD           & Phase & \multicolumn{3}{c}{$t_{\rm exp}$} & \multicolumn{3}{c}{S/N pix$^{-1}$ (s)} & \multicolumn{3}{c}{$N_{\rm LSD}$ (\%)} \\
        & ($-$2440000)  &       &  V & Q & U & V & Q & U & V & Q & U \\
\noalign{\smallskip}
\hline
\noalign{\smallskip} 
 03 Feb 00 &11578.555 & 0.940 &  2600 & 2600 & 2600 & 410 & 410 & 380 & 0.010 & 0.010 & 0.012\\
 09 Feb 00 &11584.425 & 0.470 &  1800 & 2600 & 2600 & 300 & 320 & 240 & 0.014 & 0.013 & 0.019\\
 11 Feb 00 &11586.556 & 0.026 &  1800 & 2600 & 2600 & 180 & 280 & 360 & 0.023 & 0.016 & 0.012\\
 21 Feb 00 &11596.520 & 0.624 &  2600 & -    & -    & 330 & -   & -   & 0.012 & -     & - \\
 22 Feb 00 &11597.766 & 0.949 &  2650 & 2600 & 2600 & 380 & 380 & 380 & 0.011 & 0.012 & 0.011\\
 26 Feb 00 &11601.519 & 0.928 &  2600 & 2600 & 2600 & 280 & 310 & 330 & 0.014 & 0.014 & 0.013\\
 28 Feb 00 &11603.512 & 0.448 &  2600 & -    & -    & 270 & -   & -   & 0.017 & -     & - \\
 02 Mar 00 &11606.479 & 0.222 &  2600 & 2600 & 2600 & 290 & 280 & 240 & 0.014 & 0.015 & 0.018\\
 04 Mar 00 &11608.496 & 0.747 &  2600 & -    & -    & 390 & -   & -   & 0.011 & -     & - \\
 01 Dec 01 &12245.637 & 0.897 &  1200 & 2400 & 2400 & 210 & 370 & 310 & 0.018 & 0.011 & 0.014\\
 02 Dec 01 &12246.611 & 0.151 &  1200 & 2400 & 2400 & 240 & 330 & 340 & 0.017 & 0.012 & 0.012\\
\hline\hline\noalign{\smallskip}
\end{tabular}
\caption[]{Journal of MuSiCoS spectropolarimetric observations. Phases correspond to the midpoint of the Stokes $V$ exposure, are calculated according to the new ephemeris described in the text, and exposure times are in seconds.  Reported S/N are {\em peak} S/N per $4.5$~\kms\ pixel. $N_{\rm LSD}$ is the noise level in the associated LSD profile, per $4$~\kms\ LSD velocity bin. }
\label{tab:journal}
\end{table*}

   \begin{figure}[t]
   \centering
   \includegraphics[width=8.0cm]{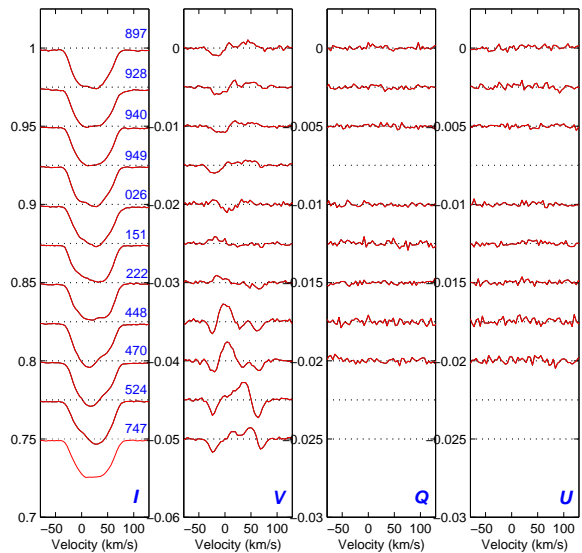}
      \caption{LSD mean Stokes $IVQU$ profiles. Rotational phases
      (according to the ephemeris given in Eq. (2) and multiplied by
      1000) are given in the left panel. Apart from a marginal detection at phase
      0.897, Stokes $Q$ and $U$ profiles are consistent with zero
      linear polarisation across the mean line. Variability of the
      Stokes $I$ profile is observed, and modulation of the
      Stokes $V$ profile shape by an apparently non-uniform Fe surface abundance
      distribution (analogous to that encountered for $\theta$~Aur by
      Wade et al. 2000a, for example) supports the view that variability of
      the mean Fe line results primarily from non-uniform surface
      abundances (see Sect. 6).}
         \label{}
   \end{figure}

 Circular polarisation (Stokes $V$) and linear polarisation (Stokes $Q$ and $U$) 
 spectra of 36 Lyn were obtained
 during  Feb/Mar 2000 and Dec 2001 using the MuSiCoS spectropolarimeter
 mounted on the 2 metre Bernard Lyot telescope at Pic du Midi
 observatory. 
 The spectrograph and polarimeter module are described
 in detail by Baudrand \& B\"ohm (1992) and by Donati et al. (1999)
 respectively. The standard instrumental configuration allows for the
 acquisition of circular or linear polarisation spectra with a
 resolving power of about 35,000 throughout the range 4500-6600~\AA.

 A complete Stokes $V$ circular polarisation observation consists of a series of
 4 subexposures between which the polarimeter quarter-wave plate is
 rotated back and forth between position angles of
 $-45\degr$ and $+45\degr$. This procedure results in exchanging the
 orthogonally polarised beams throughout the entire instrument, with
 the aim of reducing systematic errors in polarisation measurements of
 sharp-lined stars to below a level of about 0.01\% (Wade et
 al. 2000a). The procedure for recording Stokes $Q$ and $U$ linear polarisation observations is
 similar, except that the entire instrument is rotated (rather than the quarter-wave plate,
 which is removed from the beam for linear polarisation observations). Further details
 about the observing procedure are provided by Wade et al. (2000a). 

 In total 11 Stokes $V$, 8 Stokes $Q$ and 8 Stokes $U$ spectra of 36 Lyn were obtained, 
 with peak signal-to-noise ratios (S/N) of typically
 300:1 per pixel in Stokes $V$ and 325:1 in Stokes $Q$/$U$. In addition, during
 each run observations of various magnetic and non-magnetic standard stars were obtained
 within the context of other observing programmes (e.g. Shorlin et al. 2002, Kochukhov et al. 2004) which serve
 to verify the nominal operation of the instrument.

 The log of spectropolarimetric observations of 36 Lyn is shown in Table 2.

\subsection{IUE spectra}

The IUE spectra employed in this study were the NEWSIPS-extracted,
large aperture high-dispersion files from the Short Wavelength Prime
(SWP) camera. These were downloaded from the
MAST data archive. The data used were all
24 IUE SWP/large-aperture spectra for 36 Lyn, obtained during 3
epochs in 1985, 1987 and 1988. The data are described in more detail
by Smith \& Groote (2001).

\subsection{Balmer-line Zeeman analyser polarimetric observations}

Three measurements of the longitudinal magnetic field of 36 Lyn were obtained in 1992 using the University of Western Ontario (UWO) photoelectric
 polarimeter mounted on the 1.2 m telescope at the UWO Elginfield
 observatory. The polarimeter was employed as a Balmer-line Zeeman
 analyser to measure the fractional circular polarisation in the wings
 of H$\beta$ at $\pm 5.0$~\AA\ from line centre. The longitudinal magnetic field was inferred from measured fractional circularly polarised flux using a conversion factor of 17500 G/\%, as determined from scans of the H$\beta$ profile of 36 Lyn. The instrument and observing technique are described in detail by Landstreet (1982). The measurements are reported in Table 3.

\section{Least-Squares Deconvolution and longitudinal magnetic field}

 We begin by applying the Least-Squares Deconvolution (LSD) multiline
 analysis procedure (Donati et al.\ 1997) to each of the MuSiCoS
 polarised spectra in order to obtain mean Stokes $IQUV$ profile sets
 (e.g. Wade et al. 2000a). LSD mean profiles were extracted using
 line masks derived from {\tt VALD} line lists, corresponding 
 to $T_{\rm eff}=13500$~K, $\log g=4.0$ (see Sect. 5), a minimum line depth of 10\%
 and an abundance table specific to 36 Lyn (using abundances
 derived in Sect. 6). More
 information about the extraction of LSD profiles and the construction
 of LSD line masks is provided by Wade et al. (2000b) and Shorlin et
 al. (2002).

 The LSD profiles (extracted for lines of Fe) are shown in Fig. 1. The
 Stokes $V$ profiles show complex structure, are clearly detected at
 all phases, and are strongly modulated as the star
 rotates. Remarkably, Stokes $Q$ and $U$ are never confidently
 detected, notwithstanding a typical LSD noise level below 0.015\%. The Stokes $I$ profiles also show variability, which we
 interpret as rotational modulation of a non-uniform surface abundance
 distribution of Fe. This will be discussed in further detail in
 Sect. 7.
 
The longitudinal magnetic field \bz\ and its formal
uncertainty $\sigma_B$ were inferred from each of the extracted LSD
Stokes $I/V$ profile sets by numerical integration over velocity{\rm , in the manner described by Wade et al. (2000b).}
 
The 11 new LSD longitudinal field measurements
have typical uncertainties $\sigma_B\sim 60$~G, and range from -790 G
to +876 G. The inferred values of the longitudinal field of 36 Lyn are
reported in Table 3.




 
\begin{table}
\begin{tabular}{ccr}
\hline
HJD           & Phase & $\langle B_z\rangle$ \\
($-$2450000)  &       & (G)\\
\noalign{\smallskip}
\hline
\noalign{\smallskip} 
 2448739.621 &0.622  & $  -90 \pm    200$ \\
 2448744.609 &0.922  & $ -790 \pm    320$ \\
 2448754.676 &0.548  & $  660 \pm    260$ \\
\noalign{\smallskip}
\hline
\noalign{\smallskip} 
 2451578.555 &0.940  & $ -612 \pm    48$ \\
 2451584.425 &0.470  & $  595 \pm    66$ \\
 2451586.556 &0.026  & $  -91 \pm   109$ \\
 2451596.520 &0.624  & $   24 \pm    56$ \\
 2451597.766 &0.949  & $ -762 \pm    52$ \\
 2451601.519 &0.928  & $ -694 \pm    68$ \\
 2451603.512 &0.448  & $  662 \pm    82$ \\
 2451606.479 &0.222  & $  876 \pm    68$ \\
 2451608.496 &0.747  & $ -673 \pm    50$ \\
 2452245.637 &0.897  & $ -790 \pm    90$ \\
 2452246.611 &0.151  & $  609 \pm    87$ \\
\hline\hline\noalign{\smallskip}
\end{tabular}
\caption[]{New longitudinal magnetic field measurements of 36~Lyn
obtained from Balmer-line photopolarimetry (HJD 2448739-2448754) and from MuSiCoS LSD mean Stokes $I$ and $V$ profiles (HJD 2451578-). Phases are
calculated according to the rotational ephemeris, Eq. (1), determined
in Sect. 4. All uncertainties are $1\sigma$.}
\label{tab:bz}
\end{table}

We have combined the 14 new measurements of the longitudinal
magnetic field with 24 previously published measurements (2 reported
by Borra, Landstreet \& Thompson (1983) and 22 reported by Shore et
al. (1990), all obtained using a Balmer line Zeeman analyser at
H$\beta$) in order to determine the magnetic period. The
periodogram of the longitudinal field data, obtained using a modifed Lomb-Scargle technique, is characterised
by a strong, unique peak at $3.83495 \pm 0.00003$~days (resulting in a reduced $\chi^2$ of 2.07 for a first-order harmonic fit, and where the formal $1\sigma$ error bar is derived using the reduced $\chi^2$ statistical tables of Bevington 1969). This period is
consistent with, although more precise than, published magnetic and photometric values of the
period (e.g. Shore et al. 1990, confirmed by Adelman (2000), $P_{\rm
rot}=3.8345\pm 0.001$~days). This period has been kindly confirmed using minimum-false alarm probability methods by P. Reegen, who finds a best-fit period of 3.83495 days with a significance $\log S=3.34$. 

It should be noted that a second-order fit to the magnetic data provides a significantly lower reduced $\chi^2$ (1.31). Independent fitting of the individual datasets (the Balmer-line measurements and the LSD metallic-line measurements) demonstrates that the Balmer-line data are well-described by a first-order fit, and that the second-order contribution comes entirely from the LSD data. The shapes of the longitudinal field variation as defined by the two datasets are therefore quite clearly different. This is not too surprising - it is well documented that longitudinal field variations derived from metallic-line measurements often contain non-sinusoidal contributions (e.g. Wade et al. 2000b), likely due to the presence of non-uniform distributions of these elements across the stellar surface.




\section{Variations of the H$\alpha$ profile}


   \begin{figure}
   \centering
   \includegraphics[width=6.5cm]{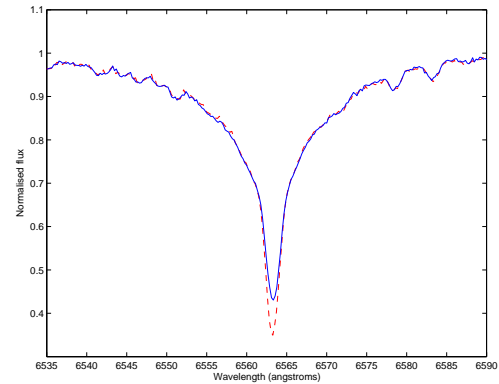} 
      \caption{Variation of the H$\alpha$ profile core of 36 Lyn due to the magnetically-confined circumstellar disk (see Sect. 4 and 7). }
         \label{}
   \end{figure}

   \begin{figure}[ht]
   \centering
   \includegraphics[width=6.5cm,angle=-90]{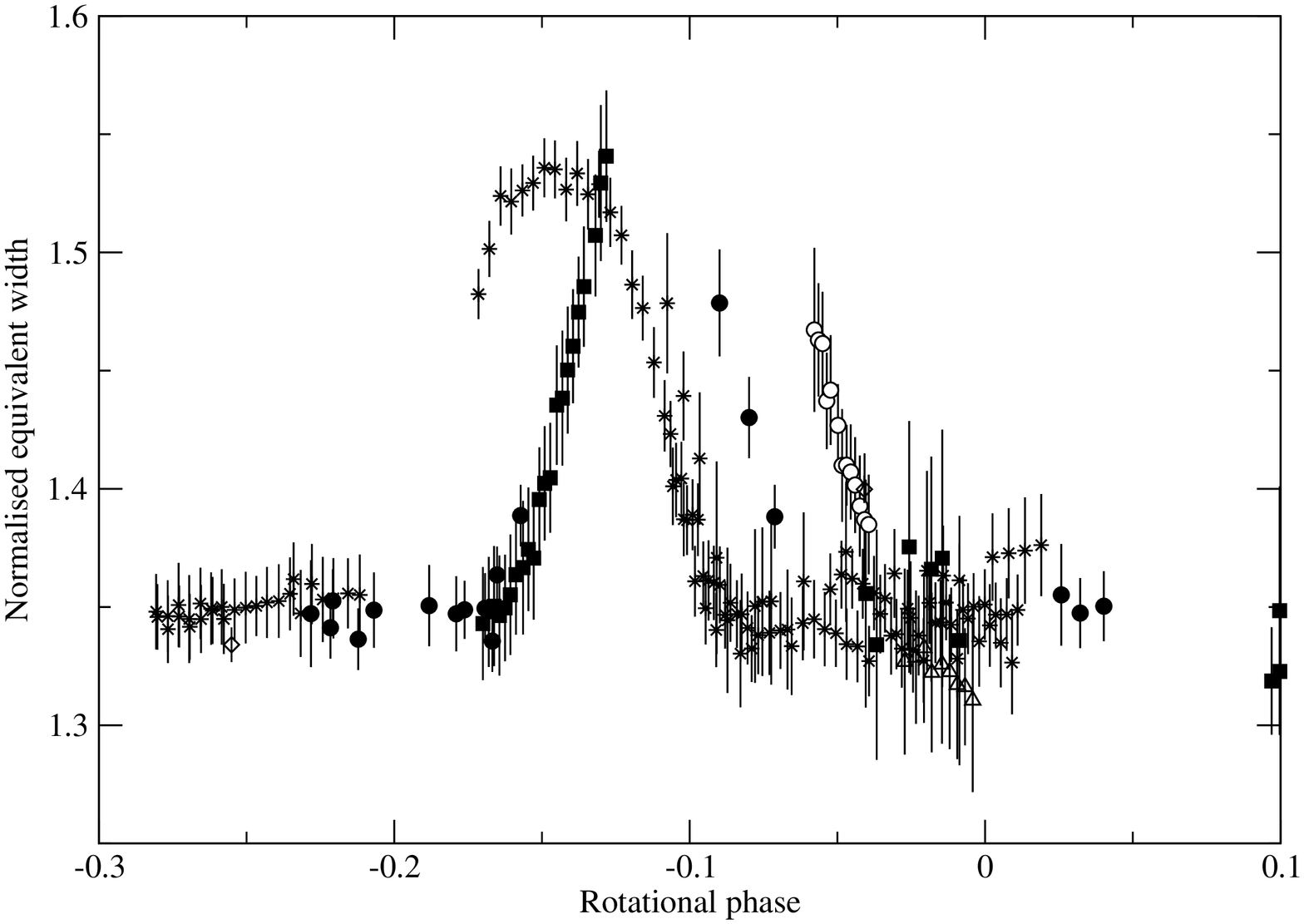} \includegraphics[width=7.5cm]{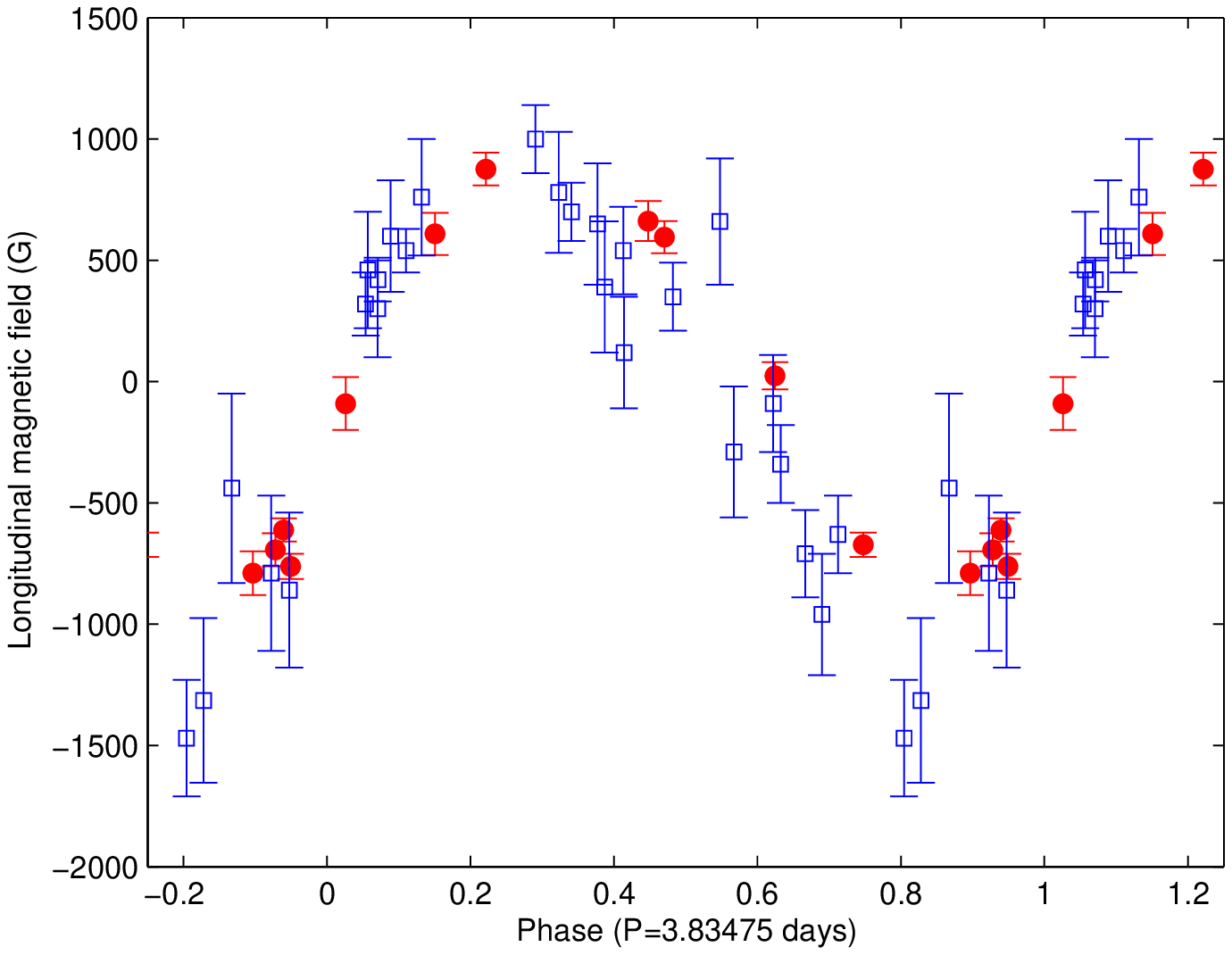} 

      \caption{{\em Upper frame -}\ Variation of the H$\alpha$ profile core equivalent width and longitudinal field of 36 Lyn. Detail of primary occultation, measurements from all spectra, phased according to the adopted {\em magnetic rotational period of 3.83495 days}. Note the clear and systematic offset of measurements obtained at different epochs (represented by different symbols).\ {\em Lower frame -}\ Longitudinal magnetic field measurements phased according to the {\em equivalent width period 3.83475 days}. Note the apparent offset between the older (H$\beta$, open squares) and more recent (MuSiCoS, filled circles) longitudinal field measurements.}
         \label{}
   \end{figure}

Upon examination of the optical spectra, it was immediately noted that
the core depth of the H$\alpha$ line of 36 Lyn is strongly
variable. Variability was first reported by Takada-Hidai \&
Aikman (1989), who attributed it to occultation of the stellar disc by
the magnetically-confined circumstellar plasma. These authors obtained several observations of H$\alpha$, and noted that the core occasionally weakens, possibly accompanied by a depression of the inner wings. We confirm the
variability, which is clearly intrinsic to H$\alpha$, and not (for example)
the result of blending with strongly-variable metallic lines (the metallic-line variability of
36 Lyn is too weak, and no suitably-located lines exist in any case) or due
to continuum registration errors (the continuum near H$\alpha$ in our spectra
is uniformly excellent, and shows internal and external agreement amongst the
datasets to within about 1\%). However, in contrast to the observations of Takada-Hidai \& Aikman, we observe no variability of the inner wings, and we observe a core which occasionally strengthens.

We have measured equivalent widths of the core of
H$\alpha$ in all of our spectra, in the range 6561.0 to 6565.5~\AA,
after local reregistering of the spectra to further reduce systematic
differences in continuum normalisation\footnote{The H$\alpha$ equivalent width measurements
are reported in Table 1, available only in electronic form from the CDS.}. Uncertainties associated
with the equivalent widths were also calculated by adopting the
inverse continuum signal-to-noise ratio as an approximate standard error bar
associated with each spectrum pixel, and propagating these errors
through the equivalent width calculation. We find that the measured H$\alpha$
equivalent width varies rather rapidly as a function of phase,
with significant variations of about 15\% (see Fig. 2).

When the H$\alpha$ equivalent width measurements are folded according
to the magnetic period determined in Sect. 3, the various epochs are found
to be {offset in phase relative to one another by up to 0.06
cycles}, as shown in Fig. 3 (upper frame). If we interpret these offsets to be the result of an error in
the period determined from the magnetic measurements, they imply a
best-fit equivalent width period (determined using Phase Dispersion Minimisation; Stellingwerf 1978) of $3.83475\pm 0.00002$ days. 
When phased according to this period, the equivalent width
data show a clear, coherent variation, with a sharp maximum near phase
0.3, as well as a much weaker increase about one-half cycle
later (Fig. 4). We interpret these 
variations in the context described by Shore, Brown \& Sonneborn (1987). In this context, the increase in H$\alpha$ absorption results from occultation of the stellar disc by plasma which is confined magnetically near the magnetic equator, and forced to co-rotate with the star. The (neutral) H gas is coupled to the ionised gas as a result of frequent reionisations, as well as by collisions. This interpretation is supported by the presence of two peaks, separated in phase by approximately 0.5 cycles, and approximately coincident with magnetic crossover phases (when $\langle B_z\rangle\sim 0$). It is further supported by observed systematic changes in the radial velocity (RV) of the H$\alpha$ core coincident with the occultation phases. As would be expected, the core RV is initially blueshifted (with respect to the systemic RV) at the beginning of the occultation, and becomes redshifted as the occultation ends (this is illustrated in Fig. 5).

The intensities of the two occultation peaks are clearly unequal. This implies that the column depth of neutral H occulting the disc differs substantially at phases 0.0/0.5, i.e. that the occulting material is not distributed uniformly within or near the magnetic equatorial plane. The accumulation of gas overdensity at the intersections of the magnetic and rotational equators (i.e. at the nodal points) of magnetic Bp stars, and differences in the column density of these ``clouds'', has been documented for the He-strong star $\sigma$ Ori E (e.g. Landstreet \& Borra 1979), for example.

The physical conditions, structure and dynamics of the magnetically-confined wind will be discussed in more detail in Paper II.

  \begin{figure*}[t]
   \centering
\includegraphics[width=7.5cm,angle=-90]{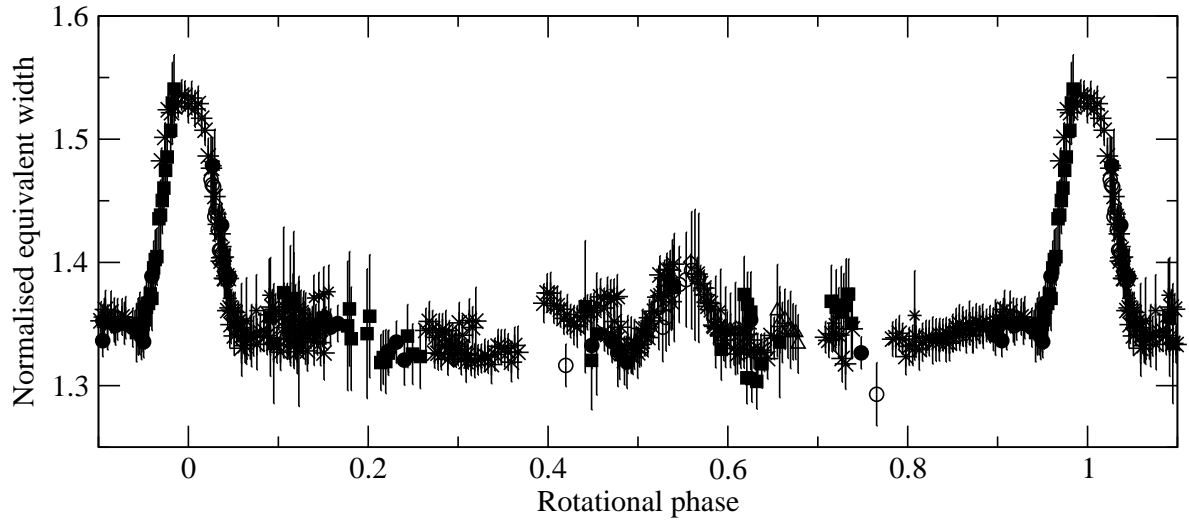} 
      \caption{Variation of the H$\alpha$ profile core equivalent width of 36 Lyn, with measurements phased according to the equivalent width period 3.83475 d. Note the significant difference in amplitude of the $\phi\simeq 0.0$ and $\phi\simeq 0.5$ peaks. {\rm Error bars shown correspond to $2\sigma$ confidence.}}
         \label{}
   \end{figure*}

{The period deduced from the H$\alpha$ equivalent
width data is not formally consistent with that determined using the magnetic data.} 
{\rm However, when the magnetic data are phased according to the
H$\alpha$ period, an acceptable phase variation is achieved. Although the new magnetic data show a small apparent offset in phase with respect to the archival data (Fig. 3, lower frame), this offset could easily result from well-known systematic differences in the shapes of field variations diagnosed from metallic vs. Balmer lines.}

{\rm We therefore adopt the H$\alpha$ period as the rotational period of 36 Lyn.} Hereinafter, all data have been phased according to the ephemeris:

\begin{equation}
{\rm JD} = (2443000.451\pm 0.03) + (3.83475\pm 0.00002)\cdot {\rm E},
\end{equation}

\noindent where the zero-point corresponds to the epoch of maximum H$\alpha$ core absorption. 

   \begin{figure}[h]
   \centering
\includegraphics[width=7.3cm,angle=-90]{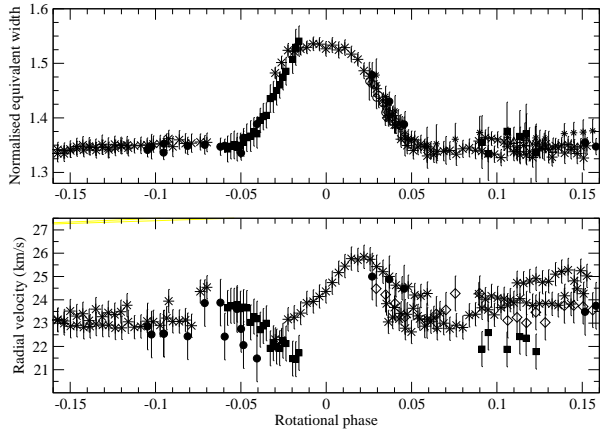} 
      \caption{Variation of the equivalent width and radial velocity of the H$\alpha$ profile core of 36 Lyn during the primary occultation at phase 0.0. Measurements have been phased according to the equivalent width period 3.83475 d. {\rm Error bars shown correspond to $2\sigma$ confidence.}}
         \label{}
   \end{figure}

\section{Fundamental characteristics of 36 Lyn}

\subsection{Fundamental parameters and evolutionary status}

We begin by redetermining the effective temperature and surface gravity
of 36 Lyn using published Geneva and $uvby\beta$ photometry, the
observed spectral energy distributions at optical and UV wavelengths,
and the H$\alpha$ and H$\beta$ profiles in our spectroscopic data.

Geneva photometry (from the Geneva photometric database, {\tt
http://obswww.unige.ch/gcpd/ph13.html}, Burki 2002) coupled with the
calibration of North \& Nicolet (1990) give $T_{\rm eff}=13650$~K and
$\log g=3.75$. 
Str\"omgren photometry (Hauck \& Mermilliod 1998)
coupled with the calibration of Moon \& Dworetsky (1985) give $T_{\rm
eff}=13740$ and $\log g=3.65$.

Using theoretical flux spectra, we have fit the observed optical
(Adelman \& Pyper 1983) and UV (Jamar et al. 1976) spectrophotometry
of 36 Lyn.  Using a modified version of the ATLAS9 model atmosphere code (Kurucz 1993) and our photospheric abundance determinations
(Sect. 6), we have performed this calculation { for a grid of temperatures and gravities bracketing the photometric values, and various metallicities ranging from $1-10\times$ solar. We have offset} magnetic broadening in the atmosphere
model by assuming a 2~\kms\ microturbulence. { In Fig. 6 we show the observed
energy distribution along with selected models.} The best model fit to the
energy distribution of 36 Lyn ($T_{\rm eff}=13300$~K, $\log g=4.0$,
[M]=0.5) is shown in Fig. 6 (upper panel), { as well as fits} for ($T_{\rm eff}=13600$~K, $\log g=3.7$, [M]=0.0) and ($T_{\rm eff}=13000$~K, $\log g=4.0$, [M]=1.0). For the UV fluxes, we used TD1 data which was converted from
flux into monochromatic magnitudes. We then reduced to 5000~\AA\ using
$m$(5000~\AA)=5.37 (a value very close to the known $m_V$=5.32).

Finally, using ATLAS9 model Balmer line profiles for solar and
enhanced metallicities, we have fit the observed { wings} of the H$\alpha$
and H$\beta$ { profiles} and obtain a best-fit for $T_{\rm eff}=13600$~K and $\log
g=3.7$ (Fig. 6, lower frames).

All three models provide a satisfactory fit to the observed flux
distribution in the optical region, while they differ in the UV
region. {\rm As $E(B-V)=0.003$ for 36 Lyn (see, for example, Lucke \cite{L78}), this cannot be explained as due to unaccounted-for reddening.} The model with the highest metallicity, 13000g40p10k2,
provides the better fit below $\lambda$=1800~\AA\,
(1/$\lambda$(microns)=5.5), while the moderate metallicity model,
13300g40p05k2, fits better in all spectral regions longer than 1800~\AA. The higher temperature model does not fit the UV flux
distribution, but it provides the best fit to the observed hydrogen
line profiles. {These { systematic and significant} discrepancies probably show { true} limitations of { ATLAS9} model atmospheres in describing the photospheric structure of 36 Lyn.} 

As a compromise, we have adopted $T_{\rm eff}=13300\pm 300$~K for
36 Lyn, and have chosen a model with $T_{\rm eff}$=13300 K, $\log
g$=4.0, and a logarithmic metallicity with respect to the sun of 0.5 for our abundance analysis. { These values have been selected giving approximately equal weight to the energy distribution and Balmer lines, although we point out that they are rather arbitrary - it is really impossible to confidently select any model within the range $13000\leq T_{\rm eff}\leq 13600$~K, $3.7\leq\log g\leq 4.0$ or $0.0\leq$[M]$\leq 1.0$.}

\begin{figure*}[t]
\centerline{\includegraphics[width=120mm]{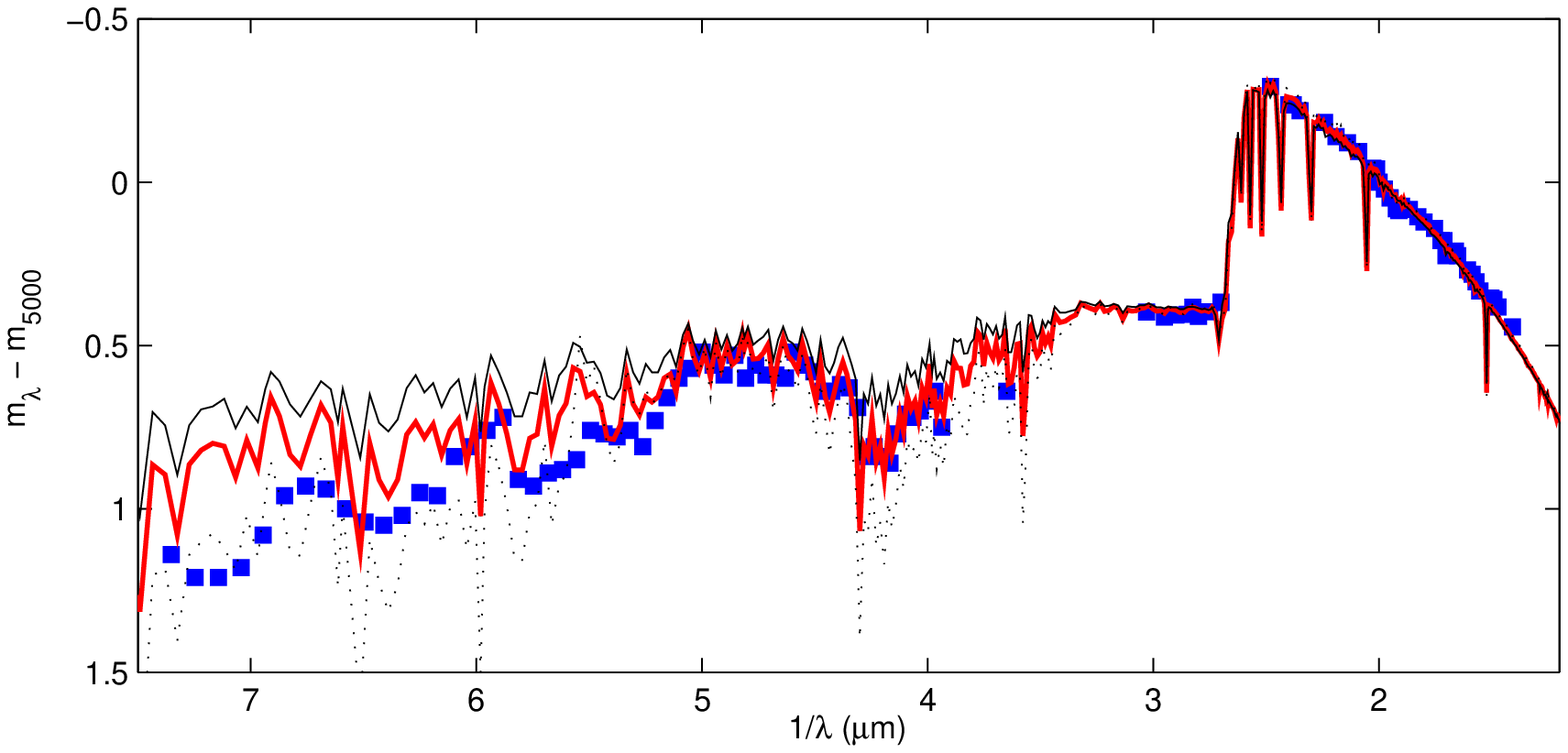}}
\vspace{0.25cm}

\centerline{\includegraphics[width=80mm]{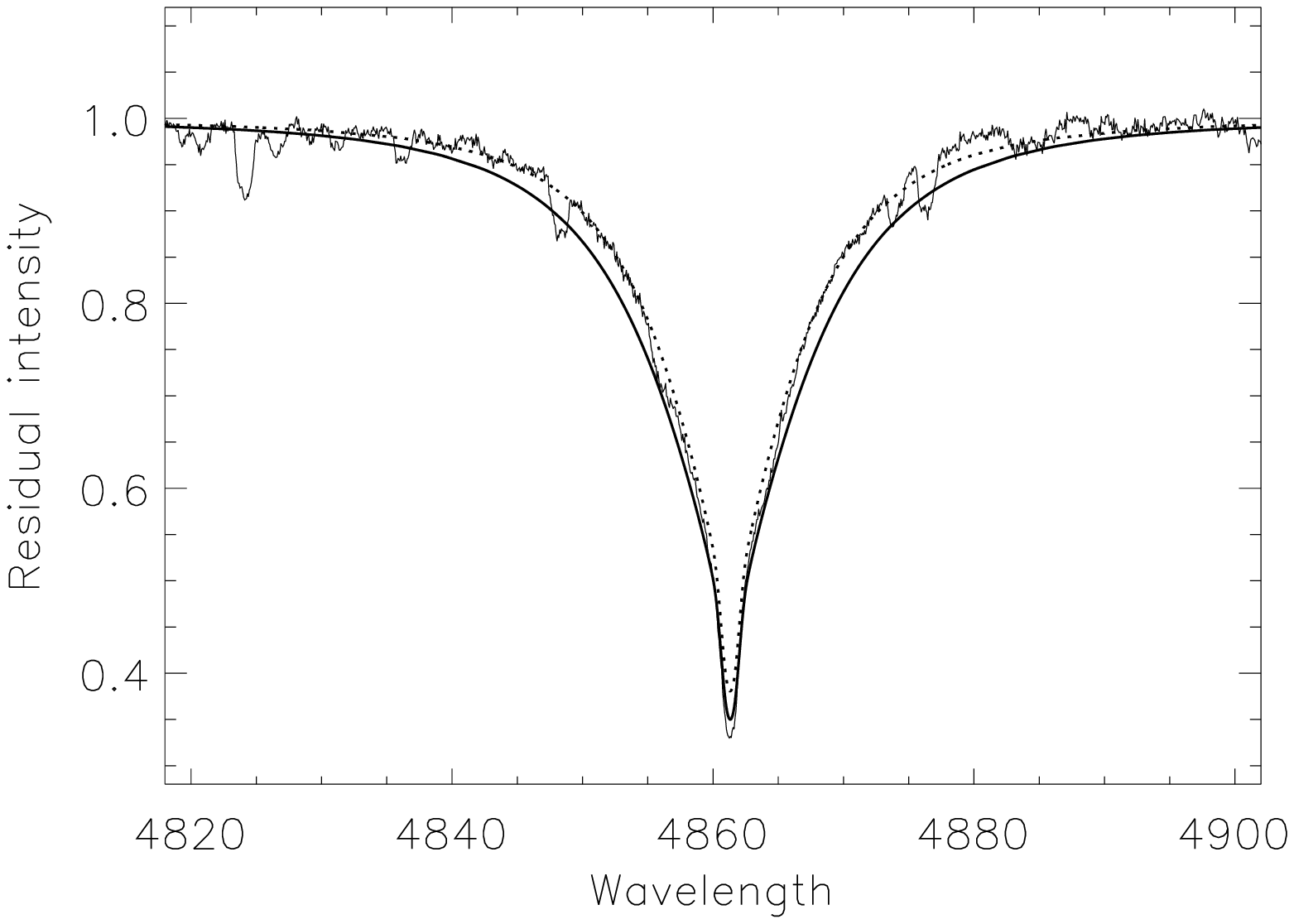}\includegraphics[width=80mm]{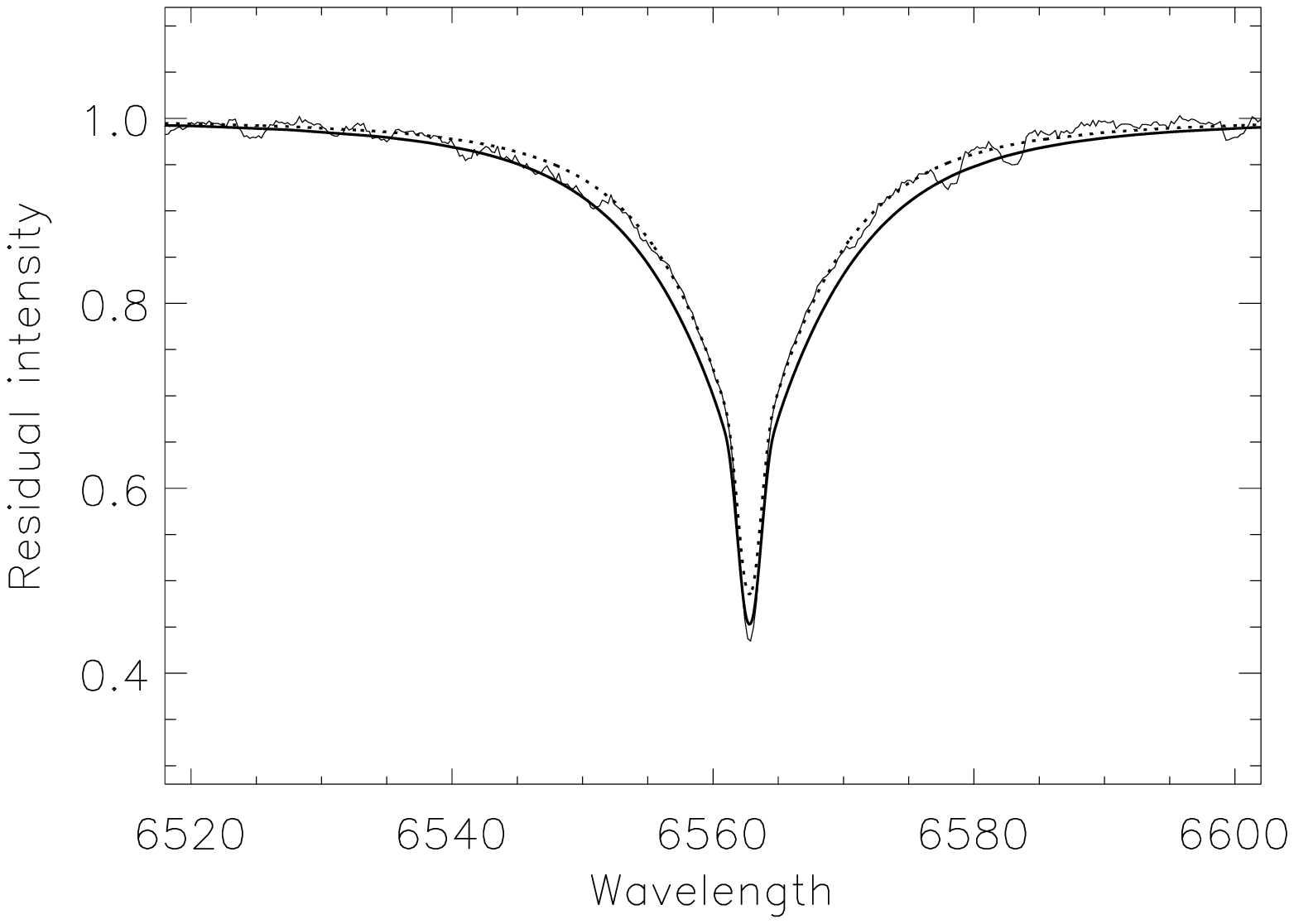}}
\caption{A comparison between the observed flux distribution in 36 Lyn (filled squares) and the theoretical
distributions calculated with three { selected} atmospheric models: 13000g40p10k2 (dotted line), 13300g40p05k2 (thick line, and the adopted model), and
 13600g37p00k2 (thin line). A comparison between the observed H$\beta$ (left panel) and H$\alpha$ (right panel) line profiles and synthesized with 13300g40p05k2 (full thick line) and 13600g37p00k2 (dotted line) models. 
}
\label{H_prof}
\end{figure*}

Now, using the derived temperature, along with the apparent visual
magnitude ($m_V=5.3$), absolute magnitude ($M_v=-0.8$) and Hipparcos
distance ($d = 162 \pm 15$~pc) reported by Gomez et al. (1998), we
obtain the bolometric magnitude of 36 Lyn $M_{\rm bol}=-1.65\pm 0.4$
(using a bolometric correction $BC=-0.85$ determined using the
interpolation formula of Balona 1994). The luminosity is therefore
$\log L/L_\odot=2.54\pm 0.16$, and the stellar radius $R=3.4\pm
0.7~R_\odot$. From the $\log T - \log L/L_\odot$ HR diagram position
of 36 Lyn (illustrated in Fig. 7), we obtain the stellar mass $M=4.0\pm
0.2 M_\odot$ and age $\log t=8.0$ ($t=79-110$~Myr) based on the
model evolutionary calculations for solar metallicity of Schaller et
al. (1992). The combination of $R$ and $M$ give a logarithmic surface
gravity of $\log g=4.0\pm 0.2$. We point out, as elaborated by Bagnulo et al.
(2005), that the use of a bolometric corrections 
derived for normal stars, coupled with comparisons with solar metallicity
evolutionary tracks, and uncertainties regarding the application of the Lutz-Kelker correction,
likely underestimates the luminosity and age uncertainties.

The measured projected rotational velocity $v\sin i=48\pm 3$~\kms,
obtained using individual metallic line profiles (see Sect. 6), and
rotational period of $3.83475\pm 0.00002$ days imply a projected radius
(assuming rigid rotation) of $R=(3.64\pm 0.23)\sin i\ R_\odot$. The radius determined from the HR diagram is numerically smaller than the projected radius determined from rigid rotation (although they are consistent within the uncertainties). The combination of the two radius estimates
implies $i\geq 56\degr$, with a best-fit value of $i=90\degr$, and a
true rotational speed in the range 45-61.5~\kms.

The derived fundamental and rotational characteristics of 36~Lyn are summarised in Table 4.

 
\begin{table}[t]
\begin{tabular}{rr}
\hline
\noalign{\smallskip} 
Mass (${\cal M}_\odot$)   & $4.0\pm 0.2$\\
Radius (${\cal R}_\odot$) & $3.4\pm 0.7$\\
Age (y)                   & $0.79-1.1\times 10^8$\\
$T_{\rm eff}$ (K)         & $13300\pm 300$\\
$\log g$                  & $3.7-4.2$\\
$v\sin i$                 & $48\pm 3$\\
Period                    & $3.83475\pm 0.00002$~d\\
$\sin i$                  &0.83-1.00\\
\hline\hline\noalign{\smallskip}
\end{tabular}
\caption[]{Fundamental characteristics derived for 36 Lyn}
\label{tab:bz}
\end{table}

\begin{figure}
   \centering
   \includegraphics[width=8.5cm]{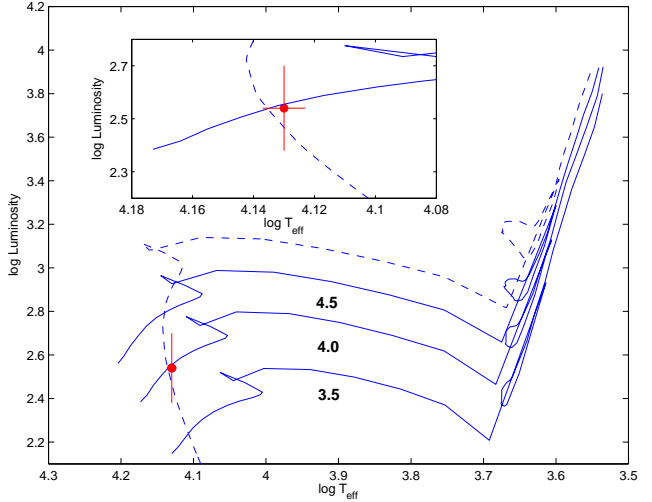}
      \caption{HR diagram position of 36 Lyn. The filled circle represents the derived position of 36 Lyn, solid curves are evolutionary tracks for 3.5, 4.0 and 4.5 $M_\odot$ for solar metallicity (Schaller et al. 1992), and the dashed curve is the associated isochrone for $10^8$ y.}
         \label{}
   \end{figure}

\subsection{Binarity}

There is no indication of binarity in the literature, and the Hipparcos mission classifies 36 Lyn as
single. The Stokes $I$ LSD profiles are constant in
mean radial velocity at $+22\pm 2$~\kms. The H$\alpha$ profiles, outside of occultation phases, give $+23\pm 1$~\kms.
Stickland \& Weatherby (1984) find $v_{\rm rad}=+26\pm 1$~\kms, whereas the Bright Star Catalogue (Hoffleit et al. 1995) gives +21~\kms, Aikman (1976) gives $+21.9$~\kms, Abt (1970) gives $+26.4$, and Takada-Hidai \& Aikman (1989; measured from H$\alpha$) give $+29.7\pm 1.6$~\kms. Although there exists some diversity among these measurements, this is expected given the metallic-line RV variations (due to surface features - see Sect. 6.3) and the H$\alpha$ RV variations (due to the circumstellar gas - see Sect. 4 and Paper II). Therefore we find no strong evidence for binarity.

\subsection{Radio and X-ray emission properties}

36 Lyn was observed at 6 cm using the VLA by Linsky et al. (1992), who
found an upper limit ($3\sigma$) of $0.62$ mJy. More recently, it has
been detected (at phase 0.73 according to our ephemeris) with a flux
of $0.45\pm 0.05$ mJy at 3.6 cm (S. Drake, private communication).

Berghofer et al. (1996) found that ROSAT obtained no detection of
X-rays at the position of 36 Lyn, reporting $\log L_x < 29.84$.

\subsection{Magnetic field geometry}




\subsubsection{Longitudinal field variation}

For a tilted, centred dipole of polar strength $B_{\rm p}$, the variation of the longitudinal magnetic field $\langle B_z\rangle$ with
rotational phase $\phi$ is given by Preston's (1967) well-known
relation

\begin{equation}
\langle B_z\rangle = B_{\rm p} {{15 + u}\over{20(3-u)}} (\cos\beta\cos i +
\sin\beta\sin i\cos 2\pi(\phi-\phi_0)),
\end{equation}

\noindent where $u$ denotes the limb darkening parameter and $\phi_0$
the phase of longitudinal field maximum (equal to $u=0.4$ and
approximately $\phi_0=0.25$ in the particular case of 36 Lyn). The inclination and
obliquity angles $i$ and $\beta$ are related by

\begin{equation}
\tan\beta={{1-r}\over{1+r}} \cot i,
\end{equation}

\noindent where $r=-0.83\pm 0.1$ is the ratio of the longitudinal
field extrema according to the best-fit sine curve\footnote{Fig. 3 suggests that
there may exist a systematic difference, at the level of a few hundred G, of the longitudinal
field variation near magnetic minimum as diagnosed using metal line versis Balmer line measurements. Such differences
frequently exist, are poorly understood, and may be related to the presence of nonuniform surface
abundance distributions. In the case of 36 Lyn, sine fits to the entire dataset or to only those
measurements obtained from H$\beta$ provide $r$ values that are identical to within the uncertainties.}. 

From $r\simeq -1.0$ we obtain straightforwardly that either $i$ or
$\beta$ is close to 90$\degr$, a result consistent with the rigid
rotation model developed considering the stellar radius, period and
$v\sin i$.

For our best-fit inclination $i=90\degr$, $\beta$ cannot be determined from
Eq. (3) and Eq. (2) reduces to

\begin{equation}
\langle B_z\rangle^{max} =  B_{\rm p} {{15 + u}\over{20(3-u)}} \sin \beta.
\end{equation}

In other words, the magnetic obliquity $\beta$ and polar strength of the dipole $B_p$ cannot be disentangled when $i=90\degr$, and we can only determine their product $B_{\rm p}\sin \beta = 3380\pm 170$~G.

In the case of our minimum acceptable inclination $i=56\degr$, we obtain using Eq. (3) that $\beta=82\pm 5\degr$, and Eq. (2) gives $B_p=3760\pm 170$~G. Therefore, {\rm considering the full range of inclinations $56\degr-90\degr$ admitted by the data}, we infer from the longitudinal field variation that $3210\leq B_{\rm p}\leq 3930$ G.

\subsubsection{Stokes $V$ profiles}

Using the method described by Donati et al. (2001) and Wade et
al. (2005), we have compared the phase variation of the Stokes $V$
profiles with the {\sc Zeeman2} (Landstreet 1988, Wade et al. 2001) predictions of a
grid of 9000 different magnetic field models, obtained by
systematically varying $i$, $B_d$ and $\beta$ in the range 56-90$\degr$, 0 to -4000 G,
and 0 to $180\degr$, respectively. For each model, the reduced
$\chi^2$ of the observed and calculated Stokes $V$ profiles was
calculated.

The best-fit solutions
for the dipolar strength $B_{\rm d}$ and the
magnetic obliquity $\beta$, for all admissible inclinations, are
$B_{\rm d}=-1300$~G and $\beta=100$. Although both methods
recover effectively the same value of the obliquity ($\beta=80\degr$
and $\beta=100\degr$ produce identical longitudinal field variations),
the dipole strength obtained from the profile fits is abour 3 times
smaller than that determined from the longitudinal field.

We strongly suspect that these (significant) differences result from the important
modulation of the Stokes $V$ profiles by the non-uniform distribution
of Fe on the stellar surface. In fact, the best-fit model for Stokes $V$,
while reproducing the amplitudes of the weak signatures around phases
0.9-0.2 rather well, is totally unable to reproduce the stronger
Stokes signatures at phases 0.5-0.8. This failure is reflected in the very high
reduced $\chi^2$ statistic ($\chi^2/\nu\sim 10.4$) obtained for the Stokes $V$ profiles.
Moreover, the structure and
radial velocities of features apparent in the observed Stokes $V$ profiles (see Fig. 1)
are not evident in the calculated profiles. This suggests that the 
model we have used to calculate the profiles of the LSD mean profiles
is an { obvious} oversimplification of the true surface structure of 36 Lyn.
Additional attempts to reproduce this structure using higher-order
multipolar magnetic field models were unsuccessful. We therefore
conclude that the structure of the Stokes $V$ profiles is likely
due to a strongly nonuniform distribution of Fe on the surface 
of the star. This will be discussed further in Sect. 6.

Finally, we report that Stokes $Q$ and $U$ profile corresponding to the best-fit Stokes $V$ dipole model (with $|B_{\rm p}|=1300 $~G) have amplitudes similar to the typical noise in the associated LSD profiles. For the dipole model derived from the longitudinal field data (with $|B_{\rm p}|=3380$~G), the predicted linear polarisation amplitudes are several times larger than the observed Stokes $Q$ and $U$ upper limits. This is consistent with results obtained for other magnetic Ap/Bp stars by Wade et al. (2000a), and may reflect the influence of non-uniform abundance distributions, or small-scale structure of the magnetic field, on the disc-integrated polarimetric signal.

\section{Photospheric chemical abundances}

\subsection{Mean abundances}

   \begin{figure*}[t]
   \centering \includegraphics[width=8.5cm]{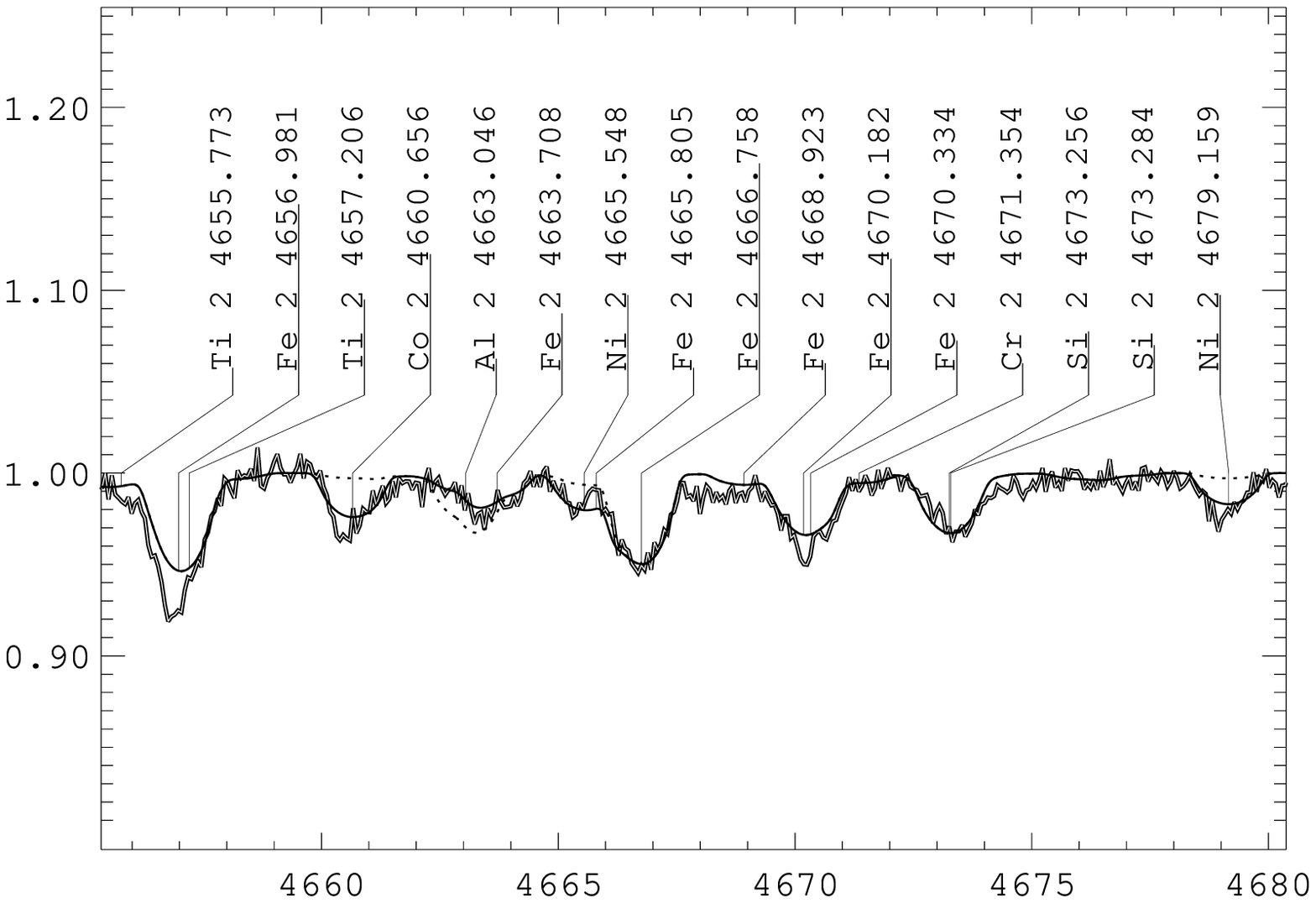} \includegraphics[width=8.5cm]{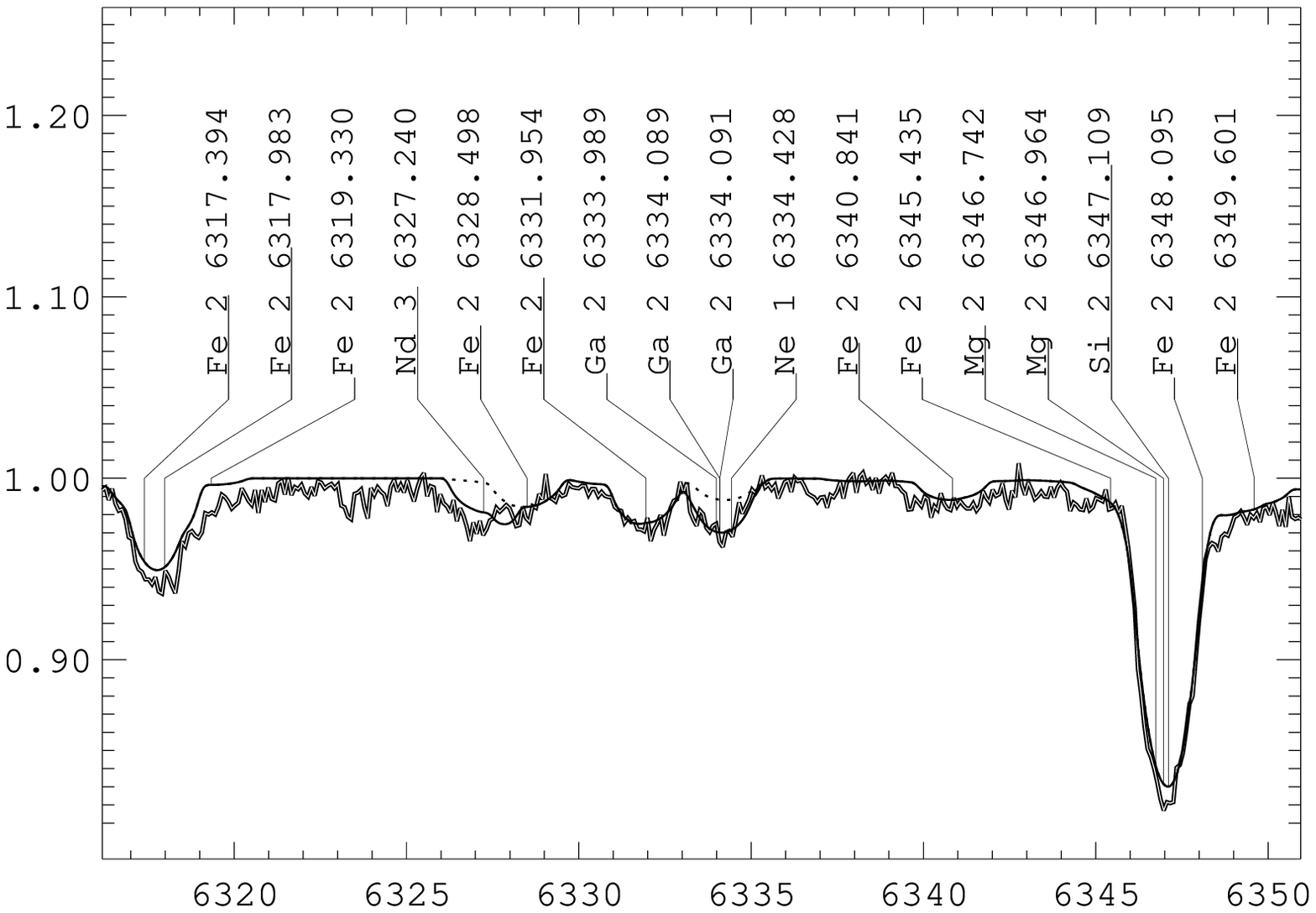}

      \caption{A comparison between the observed and synthesized spectra in the regions of Al\ii, Co\ii, 
and Ni\ii\, lines (left panel) and Si\ii, Ga\ii, and Nd\iii\, lines (right panel). Calculations with the
Al, Co, Ni, Ga and Nd abundances derived from our analysis are shown by the full line, and those with
the UV abundances (except Fe and Ti) are shown by the dotted line. The dotted line also represents the Nd\iii\,
line with a 1.0 dex reduced abundance.}
         \label{}
   \end{figure*}

Mean (phase-averaged and surface-averaged) photospheric abundances were determined from analysis of both 
optical and UV spectra of 36 Lyn.
Some previously published UV abundances were also obtained from the 
literature. All abundances are reported in Table 5.

\subsubsection{Optical abundances}
 
Synthetic spectrum calculations for the whole spectral region covered
by the optical observations (4500-6600~\AA) were performed with the synthetic
spectrum codes {\sc synth} (Piskunov 1992) and {\sc starsp}
(Tsymbal 1996). Selected regions of the spectrum were also modeled
using {\sc Zeeman2} (Landstreet 1988, Wade et al. 2001). For hydrogen 
line profiles we used the more
recent calculations of the hydrogen line opacities (Barklem at
al. 2002). We did not include the magnetic field for {\sc synth} or
{\sc starsp} synthetic
calculations; instead, we employed a pseudomicroturbulence of 2 \kms\,
to account for magnetic intensification. The influence of a dipolar 
magnetic
field of surface intensity 3 kG was employed for the {\sc Zeeman2} 
calculations.
The adopted model atmosphere is { an ATLAS9 model} with the { adopted} parameters described in Sect. 
5. We have explored the range of model parameters discussed in Sect. 5.1, derived from the photometry, energy distribution and Balmer lines, and we find that the detailed choice of model atmosphere has a negligible impact on the derived abundances. With a few
exceptions, all atomic data were extracted from the VALD database (Kupka et
al. 1999, Ryabchikova et al. 1999). Rare earth abundances (Pr and Nd)
were derived from the lines of the second ionization stage with the
transition probabilities from Cowley \& Bord (1998), Bord (2000) -
Nd~{\sc iii}, and from Bord (private communication) - Pr~{\sc iii}. Two 
lines of Ga~{\sc ii}, $\lambda\lambda$ 6334, 6419 were used
for the gallium abundance determination in the optical region. Hyperfine
and isotopic splitting (Karlsson \& Litz\'en 2000) were taken into
account. 

In order to fit the line profiles, it was necessary to determine the 
projected rotational
velocity $v\sin i$. Various estimates of $v\sin i$ exist in the 
literature, ranging from 40~\kms
(Abt \& Morrel 1995) to 60~\kms (Abt, Levato \& Grosso 2002). We find 
$v\sin i=48\pm 3$~\kms,
consistent with the recent results of Royer et al. (2002) who find 
$v\sin i=49$~\kms. 

Fig. 8 shows a part of the 36 Lyn optical spectrum and synthetic 
spectrum fits.

\begin{table}
\begin{tabular}{rrrrr}
\hline
                 & 36 Lyn Optical$^1$ & 36 Lyn UV& $\odot$\\
Element          & $\log n/n_{\rm tot}$& $\log n/n_{\rm tot}$& $\log n/n_{\rm tot}$\\
\noalign{\smallskip}
\hline
\noalign{\smallskip}
He & $-1.65\pm 0.05$&                                  & -1.05 \\
 C & -3.0 - -2.7    & -3.48$^1$                         & -3.48 \\
 O & -3.4 - -3.0    &                                  & -3.11 \\
Ne & -3.8 - -3.95   &                                  & -3.95 \\
Na & -5.30          &                                  & -5.71 \\
Mg & -4.35          & $-4.44\pm 0.22^2$                & -4.46 \\
Al & -6.30          & $-6.10\pm 0.15^2$                & -5.57 \\
Si & $-4.10\pm 0.2$ &$-4.5:\pm 0.4^1$ & -4.49 \\
   &                &$-4.70\pm 0.10^2$&\\
P  & -6.30          &                                  & -6.59 \\
S  & -4.8 - -5.0    &                                  & -4.83 \\
Cl & -5.2           &                                  & -6.54 \\
Ti & $-5.00\pm 0.25$    &                                  & -7.05 \\
Cr & $-5.00\pm 0.3$   &-5.3$^1$,         & -6.37 \\
   &                &$-5.3\pm 0.4^3$   &  \\
Mn & -5.8           &$-6.80\pm 0.15^3$                 & -6.65  \\
Fe & $-3.45\pm 0.25$    &-3.2$^1$         & -4.37 \\
   &                &$-3.90\pm 0.1^3$  &       \\
Co & $-5.00\pm 0.15$           &$-6.2\pm 0.2^3$                   & -7.12 \\
Ni & $-4.90\pm 0.15$    &$-5.7\pm 0.1^3$                   & -5.79 \\
Cu &                &$-8.1\pm 0.6^4$                   & -7.83 \\
Ga & $-6.05\pm 0.1$ &$-6.90\pm 0.55^5$                 & -9.16  \\
Pr & $-8.1\pm 0.2$  &                                  & -11.33\\
Nd & $-7.7\pm 0.2$  &                                  & -10.54\\
\hline\hline\noalign{\smallskip}
\end{tabular}
\caption[]{Mean photospheric chemical abundances of 36 Lyn. Also shown 
are the respective solar abundances. Sources: $^1$This work, $^2$Smith (1993), $^3$Smith \& Dworetsky (1993), 
$^4$Smith (1994), $^5$Smith (1996).}
\label{tab:bz}
\end{table}

\subsubsection{Ultraviolet abundances}

Ultraviolet abundances were inferred from spectrum synthesis using the 
{\sc synspec} code (Hubeny, Lanz \& Jeffery 1994), assuming a Kurucz 
(1993) ATLAS9 model atmosphere.

{
Notably, we discovered that the silicon abundances derived from Si~{\sc ii} and
Si~{\sc iii} lines differed, a problem that we ultimately traced to a
series of blends in the Si~{\sc iii} lines and which may potentially lead to erroneous abundances of Si derived for Bp stars from {IUE} data (two good, unblended Si {\sc iii}
lines are those at 1370.0~\AA\~ and 1417.2~\AA.) We were finally able to distinguish the unblended and blended Si~{\sc iii} lines by their variability, as the former show variations
in their radial velocity with phase, as reported from an optical Si line study
by Stateva (1997).}



\subsection{Mean abundances and evidence for vertical abundance non-uniformities}

In our analysis of optical spectra of 36 Lyn, we find no 
systematic differences in
the abundances required to fit strong versus weak lines of elements with
sufficiently rich line spectra. This suggests that
stratification of these elements in the atmosphere of 36 Lyn is either 
very weak or nonexistent.


For a few elements, the abundances obtained from optical lines differ from those derived from UV lines.  These differences cannot be explained by small differences in the adopted atmospheric parameters. The largest disagreements occur for Mn, Co, Ni and Ga which
appear to be less abundant from the UV analysis, while Al is opposite.
The Ga problem is well known (Lanz et al.~1993). Dworetsky
\& Smith (1993) derived abundances from the low excitation
UV Mn~{\sc ii}, Co\ii\, and Ni\ii\, lines, while our analysis is based on high
excitation lines. Therefore a difference in the UV and optical
abundances may reflect stratification of these elements, although the
discrepancy could be attributed to
unidentified blends given the relative scarcity of lines of these
elements in the optical. An examination of the oscillator strengths used in the various studies indicates that systematic errors in $gf$ values cannot explain the differences. For Co~{\sc ii} lines, neglect of hyperfine structure may be an important contributor.

Overall, we find that He is depleted relative to solar, while Fe peak
elements are strongly enhanced, by one dex or more. Abundances of two
rare earth elements (REEs, Pr and Nd) are found to be enhanced by
approximately 3 dex. Mn, Co, Ni and Ga may be stratified.
      
\subsection{Horizontal abundance non-uniformities}

  \begin{figure*}
   \centering
  \includegraphics[width=11cm]{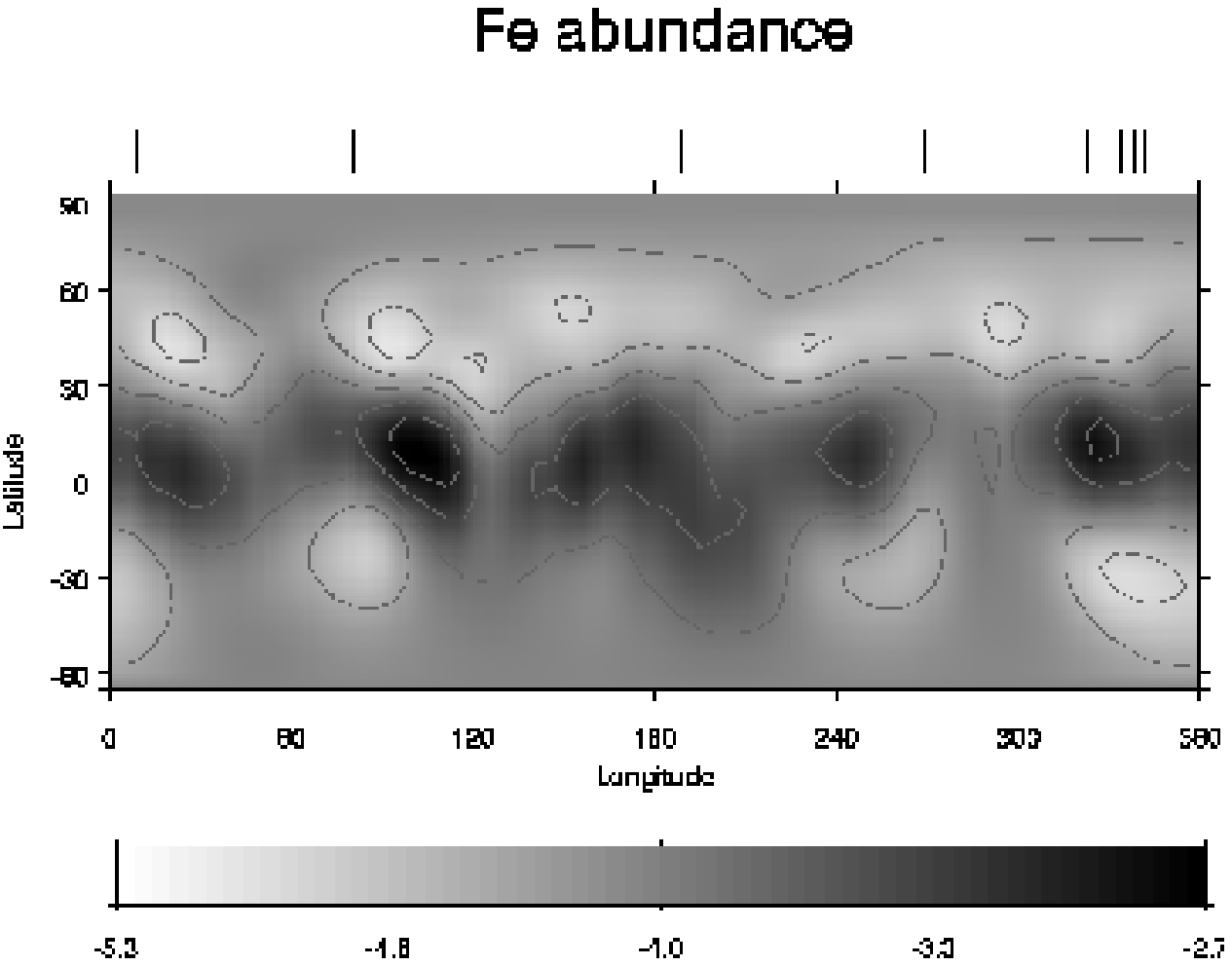}\hspace{0.5cm}
   \includegraphics[width=4.5cm]{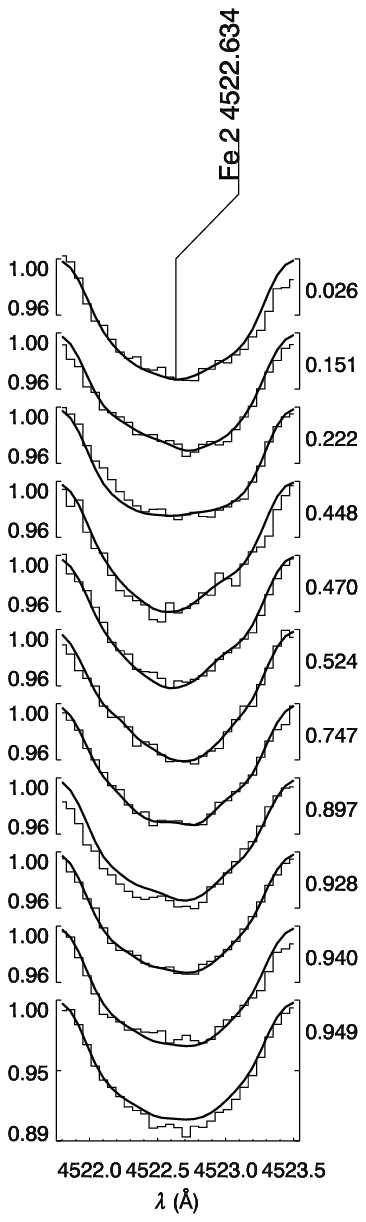}
      \caption{Fe Doppler image from Fe~{\sc ii} $\lambda$4522. {\em Right frame -}\ comparison of observed and calculated profiles. {\em Left frame -}\ Derived abundance map in rectangular projection. Tick marks indicate phases of 36\,Lyn observations used to derive the map.}
         \label{}
   \end{figure*}

The variations of lines of a number of elements in the high-resolution
optical spectrum of 36 Lyn as well as the LSD profiles are suggestive
of non-uniform horizontal distributions of their abundances. However,
some care is required in interpreting these variations: Smith \&
Groote (2001) caution that the magnetically-confined winds of Bp stars
can produce optical line profile variability due to occultation of the
stellar disc by the circumstellar material, and that in a number of
hot Bp stars a significant component of the optical metallic-line
variability can be attributed to this phenomenon.

In the spectrum of 36 Lyn, weak variability occurs in lines of He, Fe,
Si, Cr, Ti, and Mg (based on our own spectra, as well as the results
of Stateva 1997). Some of these elements show extrema of their
equivalent width phase variations which coincide with the magnetic
extrema (phases 0.25/0.75, either in phase or anti-phase, e.g. He, Fe) whereas others
show variations with extrema at crossover phases (phases 0.0/0.5, e.g. Si). We have also examined the
dependence of EW variation shape and phase of Fe lines as a function of excitation potential,
and find no { significant} dependence. Based on (1) the
variety of equivalent width variability characteristics, (2) the large range
in excitation potential of lines that undergo variability, (3) the modulation of the
shapes of Stokes $V$ profiles and (4) the apparent difference in the extrema of the
longitudinal field variations as diagnosed in metal versus H lines, we conclude 
that the optical line profile variations result substantially from
surface structure (magnetic + abundance non-uniformity), rather than
being due to the circumstellar material. This conclusion is supported by an examination of the high-resolution
line profiles. As can be seen in Figs. 1 and { 9}, Fe lines exhibit variations
in Stokes $I$ and $V$ profile shapes, consistent with Doppler
modulation by surface non-uniformities.

Stateva (1997) attempted an harmonic reconstruction of the surface
distributions of He, Si, Cr and Mg using equivalent width variations
extracted from photographic spectra of 36 Lyn, and assuming that the
variations were due entirely to surface non-uniformities. She was able
to reproduce the (relatively low signal-to-noise ratio) equivalent
width and radial velocity variations, and the resulting maps suggest
that He, Si and Cr are concentrated primarily in the plane of the
magnetic equator.

Here, we apply the more stringent test of attempting to reproduce the
detailed line profile variations using the Doppler Imaging
technique, for the particular case of Fe. Because we have only 11 high-resolution spectra, we are not attempting to derive a definitive map of the surface Fe distribution. Our primary goal is rather to demonstrate that the detailed profile variations of moderate and high-excitation Fe lines in the optical can be substantially explained by an inhomogeneous surface distribution of this element, and are not significantly affected by absorption due to the circumstellar material.   

\emph{Doppler imaging} is a technique which allows the inversion of observed spectral line profile variations
to produce a two-dimensional map of the abundance distribution at the surface of a star. An inhomogeneity on the stellar surface leads to 
distortions in the spectral line profiles.
These distortions appear on the blue side of the line profile as the spot comes into sight on the visible
hemisphere of the star,
travel across the spectral line and disappear on the red wing as the spot rotates out 
of sight. The longitudinal position of the inhomogeneity is obtained directly from the velocity position of the distortion within the profile, whereas its latitude has to be deduced from spectral time series.     
In INVERS12, the Doppler Imaging code we applied, developed by N. Piskunov and refined by O. Kochukhov (Kochukhov et al. 2004b),  
the equation of radiative transfer is solved in a way that
allows the calculation of maps of several chemical elements simultaneously
and self-consistently.
In this program the precalculated local line profile tables used in former versions 
of the code are replaced by the calculation of specific intensities for each visible
surface element on each iteration.


For 36 Lyn, we assumed various inclination angles between $56\degr$ and $90\degr$ to perform the Doppler Imaging. The best fit to the profiles was found for $i=60\degr$, although the dependence of the fit quality on $i$ was rather weak (and in particular this dependence cannot be used to constrain the inclination). We derived the Fe abundance on the surface of the star using several Fe~{\sc ii} lines, including $\lambda 4520, \lambda 4522, \lambda 5018, \lambda 5534$ and  $\lambda 5780$, each of which produced a similar Fe distribution. In Fig. 9 we show the Fe distribution, in the rotational frame of reference, reconstructed from Fe\,{\sc ii} $\lambda$4522.6 ($\chi_e=2.8$~eV). The distribution is characterised by a very patchy ring-like structure between -30\degr and 
+30\degr\ in stellar latitude, with a total contrast of somewhat more than 3 dex. Due to the uncertainty in the magnetic field geometry, it is difficult to conclude whether any obvious geometrical relationship exists between the Fe surface distribution and the magnetic field. Our success in interpreting the profile variations of the Fe~{\sc ii} lines supports our interpretation of the Fe (and other metal) line variations are due to Doppler modulation of surface structure. The details of the derived map should be interpreted conservatively, given that the map was derived using a single line with relatively coarse rotational phase coverage.

\section{Discussion and conclusion}

We have performed a thorough reevaluation and analysis of the physical and spectroscopic characteristics of the B8p He-weak star 36 Lyncis. Based on measurements of the longitudinal magnetic field, as well as measurements of the core equivalent width of the H$\alpha$ line, we have redetermined the rotational period $P_{\rm rot}=3.83475\pm 0.00003$ days. An examination of the variability of the H$\alpha$ core equivalent width reveals strong changes coincident with magnetic crossover phases, which we interpret as due to occultation of the stellar disc by magnetically-confined circumstellar gas. We have determined the effective temperature, radius, mass, age, projected and true rotational velocities, the rotation axis inclination, and we have summarised evidence for binarity, radio and X-ray emission. 

{\rm We have found { systematic and significant} discrepancies between the temperature and gravity determined from the optical and UV energy distribution and from the Balmer lines, { concluding that they probably show true limitations of the ATLAS9 model atmospheres in describing the real photospheric structure of 36 Lyn.} The ATLAS9 models we have employed are calculated using scaled solar metallicities, while 
individual abundances in
the atmosphere of 36 Lyn show large diversity. To check the influence of 
anomalous abundances
on the model atmosphere structure, a few model atmospheres were calculated with {\sc LLmodels} model atmosphere code (Shulyak et al. 
\cite{LL04}) which takes
into account the contribution of individual line 
opacities to the total continuum opacity through direct synthetic 
calculations. This code allows us to examine the influence of the particular abundance table of 36 Lyn on the model atmosphere structure. The flux distribution for an {\sc LLmodels} model with $T_{\rm eff}$=13300, $\log g=4.0$ and the abundances of 36 Lyn actually provides a somewhat better simultaneous fit
to both the energy distribution and Balmer lines. We conclude that a model atmosphere with individualised abundances 
improves the overall fit to the observables, { although discrepancies are still evident} (a surface gravity below 4.0 is still
required to fit the H lines). { Possibly we are seeing the influence of peculiar and possibly stratified abundances on the photospheric temperature and pressure structure - effects which are not included in {\sc LLmodels}.}

We have modeled Least-Squares Deconvolved Stokes $V$ profiles and the longitudinal field variation to constrain the magnetic field geometry. We have analysed both optical and UV spectra and, using spectrum synthesis techniques, we have determined the abundances of 21 elements. We find no evidence for vertical stratification of most elements, although for a few elements (e.g. Mn, Co, Ni, Ga) we have derived abundances in the optical which differ from those in the UV, suggesting that these elements may be stratified. {\rm The influence of using an {\sc LLmodels} individualised abundance model atmosphere on inferred abundances is small ($\pm$0.1 dex) for 
all elements except He, for which the abundance is decreased by 0.4 dex. On the other hand, using the individualised abundance model 
decreases by 0.15 dex the difference
between the optical and UV Ga abundances (Ga optical abundance is decreased by 0.1 dex and Ga 
UV abundance increased by 0.05 dex).}

We find convincing evidence exists for nonuniform surface (horizontal) distributions of some elements. Using the Doppler Imaging technique, we have constructed { an approximate} surface map of the Fe abundance distribution, which is characterised by a patchy ring-like structure between $-30\degr$ and $+30\degr$, and a contrast of somewhat more than 3 dex. The results reported in this paper set the stage for a detailed study of the structure and geometry of the circumstellar material in Paper II.


\begin{acknowledgements}
CTB, JDL and GAW have been supported in part by Discovery Grants from the Natural Science and Engineering Research Council of Canada. GAW has been partially supported by the Academic Research Programme of the Department of National Defence (Canada). TAR and TL have been supported by the Austrian Science Fund (FWF-P17580N2). The authors thank P. Reegen for confirming their period calculations using minimum-fap methods. TL would like to thank O. Kochukhov for valuable discussion and providing IDL plotting routines. CTB also thanks Jim Thomson and Heide DeBonde for their help on service observing nights at DDO.
\end{acknowledgements}
\end{document}